\documentstyle[aps,amssymb,preprint]{revtex}
 1

\begin{document}
\title{\boldmath Temperature and filling dependence of the 
superconducting $\pi$-phase in the Penson-Kolb-Hubbard model}
\author{Fabrizio Dolcini\footnote{e-mail: fdolcini@athena.polito.it},
and Arianna Montorsi\footnote{e-mail:montorsi@athena.polito.it}} 
\address{Dipartimento di Fisica and Unit\`a INFM, Politecnico di Torino,
I-10129 Torino, Italy}
\date{\today}
\maketitle
\begin{abstract}
We investigate in the Hartree Fock approximation the temperature and
filling dependence of the superconducting $\pi$-phase for
the Penson-Kolb-Hubbard model. Due to the presence of the pair-hopping
term, the phase survives for repulsive values of the on-site Coulomb
interaction, exhibiting an interesting filling and temperature
dependence. The structure of the self-consistent equations peculiar to
the $\pi$-phase of the model allows to explicitly solve them for the
chemical potential. The phase diagrams are shown and discussed in
dimension $2$ and $3$. We also show that, when a next-nearest neighbours
hopping term is included, the critical temperature of the
superconducting region increases, and the corresponding range of filling
values is shifted away from half-filling. Comparison with known exact
results is also discussed.

\end{abstract}
%%%%%%%%%%%%%%%%%%%%%%%%%%%%%%%%%%%%%%%%%%%%%%%%%%%
%%%%%             S E C T I O N   1           %%%%%
%%%%%%%%%%%%%%%%%%%%%%%%%%%%%%%%%%%%%%%%%%%%%%%%%%% 
\begin{section}{Introduction}

Interest in strongly correlated electron systems and superconductivity
has motivated some attention on the wide class of Extended Hubbard
models\cite{HUB}. For such models, a certain number of exact results
has been obtained for specific values of the interaction
parameters\cite{VOL,MON}, expecially in the one-dimensional case
\cite{EKS-1}-\cite{BGLZ}. These results support in particular
the existence of a superconducting phase at $T=0$, which is worth
investigating also when relaxing some of the constraints among the
parameters that allow the integrability of the Hamiltonian.

The Extended Hubbard Hamiltonian reads 
\begin{equation}
H_{EH}=H_{Hub} + H_{X}+H_{\tilde X}+H_{V}+H_{W}+H_{Y} \quad ,
\label{HEH}
\end{equation}
where 
\begin{eqnarray*}
& H_{Hub}= -t \sum_{<{\bf i},{\bf j}> }\sum_\sigma c^{\dagger}_{{\bf i},
\sigma}c^{}_{{\bf j}, \sigma} + U \sum_{\bf i} n_{{\bf i}, \uparrow}
n_{{\bf i},\downarrow} \hspace{0.5cm} \: & \mbox{(Hubbard)} \\
& H_{X}=X \: \sum_{<{\bf i},{\bf j}> }\sum_\sigma 
(n_{{\bf i},-\sigma}
+ n_{{\bf j},-\sigma}) c^{\dagger}_{{\bf i},\sigma}
c^{}_{{\bf j}, \sigma} \hspace{0.5cm} \: & \mbox{(bond-charge)} \\
& H_{\tilde X}= {\tilde X} \: \sum_{<{\bf i},{\bf j}> } 
\sum_\sigma n_{{\bf i},-\sigma}
n_{{\bf j},-\sigma} c^{\dagger}_{{\bf i},\sigma}
c^{}_{{\bf j}, \sigma} \hspace{1.3cm} \: & 
\mbox{(correl. hopping $\leftrightarrow$ on-site number)} \\
& H_{V}={V\over 2}\sum_{<{\bf i},{\bf j}> }  
n_{\bf i} n_{\bf j} \hspace{4.4cm} \: & 
\mbox{(neighboring site charge)} \\
& H_{W}= {W\over 2} \sum_{<{\bf i},{\bf j}> } 
\sum_{\sigma,\sigma^{\prime}} c^{\dagger}_{{\bf i},\sigma} 
c^{\dagger}_{{\bf j}, \sigma^{\prime}} c^{}_{{\bf
i},{\sigma}'} c^{}_{{\bf j}, \sigma} \hspace{1.4cm} \: & 
\mbox{(exchange)}  \\
& H_{Y}= Y \, \sum_{<{\bf i},{\bf j}> }
 c^{\dagger}_{{\bf i},\uparrow} 
c^{\dagger}_{{\bf i},\downarrow}  c^{}_{{\bf
j},\downarrow} c^{}_{{\bf j}, \uparrow}  \hspace{2.8cm} 
\: &  \mbox{(pair hopping)} \quad .
\end{eqnarray*}

Here $c_{{\bf i},\sigma}^\dagger$ and $c_{{\bf i},\sigma} \,$ 
are fermionic creation and annihilation operators, the 
subscript~${\bf i}$ running over the sites 
of a $d$-dimensional lattice $\Lambda$ of $L^d$ sites, and 
$\sigma \in \{ \uparrow , \downarrow
\}$ being the spin label; the usual anticommutation rules 
$\{ c^{}_{{\bf i},{\sigma}'}, c^{}_{{\bf j},
\sigma}\} = 0 \,$, $\, \{ c^{}_{{\bf i},\sigma} , 
c^{\dagger}_{{\bf j}, \sigma'} \} 
= \delta_{{\bf i},{\bf j}}\,
\delta_{\sigma,\,\sigma'}$ hold. The symbol $< {\bf i} , \, {\bf j} >$
stands for nearest neighbors in $\Lambda$. 
Finally, $n^{}_{{\bf
i},\sigma} = c^{\dagger}_{{\bf i},\sigma} 
c^{}_{{\bf i},\sigma}$ is the electron number operator with  
spin $\sigma$ at site $\bf i$, and $n^{}_{\bf i}=n^{}_{{\bf i} 
\uparrow} + n^{}_{{\bf i} \downarrow}$.  
\\As Hubbard himself pointed out in \cite{HUB}, all the terms in
(\ref{HEH}) appear due to the second quantization formulation 
of the interaction among electrons. The specific physical
system of interest then suggests which are the most relevant ones.

For the pure Hubbard model no exact result proves the 
existence of a superconducting phase at finite values of $U$, 
and even within a mean-field scheme this is achieved only for negative 
$U$ values. On the contrary a number of interesting
exact results on the existence of a superconducting phase for appropriate
non-vanishing values of some other interaction parameters in (\ref{HEH})
has been obtained. Most of these results involve the notion of 
the states known as $\eta_{\phi}$-pairs, {\it i.e.} the states 
\begin{equation}
| \eta \rangle_{\phi} = (K^{\dagger}_{\phi})^{m} 
|  0 \rangle \quad , \quad m=1,\dots, L^d \label{ETA}
\end{equation}
with 
\begin{equation}
K^{\dagger}_{\phi} = 
\sum_{{\bf j} \in \Lambda} e^{i \mbox{\small \boldmath $\textstyle
\phi$} \cdot {\bf j}} \, c^{\dagger}_{\bf j \, \uparrow} 
c^{\dagger}_{\bf j \, \downarrow} \; = \; 
\sum_{{\bf k} \in B} 
c^{\dagger}_{\small \mbox{\boldmath $\, \phi$} - {\bf k} \, 
\downarrow} 
\, c^{\dagger}_{\, \bf k \, \uparrow} \quad , 
\label{Kphi}
\end{equation}
\noindent where {\boldmath $\phi$} is a $d$-dimensional 
vector $(\phi,\phi, \dots)$, and $B$ is the first Brillouin zone in the
reciprocal lattice. Noticeably, the states $|\eta\rangle_\phi$  
enjoy the property  of `Off Diagonal Long Range Order' ($ODLRO$),
which implies superconductivity \cite{YANG}. Therefore much 
effort has been done through the last years to 
find which are the relations among the coupling parameters in (\ref{HEH})
guaranteeing that an $| \eta \rangle_{\phi}$ is the ground state.
\\A first set of remarkable results in $1$-d was obtained for the
subclass of Hamiltonians characterized by the constraint
$X=t$. In \cite{AL-AR,SCHAD} the phase diagram $U$ vs filling for
the AAS model $X=t, \tilde{X}=W=V=Y=0$ (reported in fig.\ref{pkh_fig1})
was derived at $T=0$ : one can see a superconductive 
filling-independent region, where the $| \eta \rangle_{\phi}$ 
are degenerate ground states for any $\phi$, and a filling 
dependent zone (again superconducting because it contains at least 
$| \eta \rangle_{0} $ pairs) rising up to positive values of $U$.
Unfortunately, in contrast to the real case of superconducting
materials, the superconducting phase turns out to have a maximum at
half filling. A similar phase diagram (see fig. 1) was also obtained
in \cite{EKS-1,EKS-2} for the EKS Hamiltonian,
characterized by $X=t, \tilde{X}=0, \, Y=W=V=-1$. There the filling
independent phase is made of $| \eta \rangle_{\phi}$ with just $\phi=0$,
since $Y\neq 0$. In fact a non-vanishing pair-hopping term removes
the degeneracy of $|\eta\rangle_\phi$, $| \eta\rangle_{0}$ being
energetically favourite for $Y<0$, while $| \eta\rangle_{\pi}$ is 
favourite for $Y>0$. Moreover, as fig.1 shows, a non-vanishing $Y$
also contributes to rise up the superconducting region towards positive
values of $U$.

More recently, it has also been realized \cite{MON} that, at least in
order to obtain a filling independent superconducting ground state
region, some of the constraints on the parameters (in particular $X=t$, 
which is not very physical)
are not necessary, provided that a pair-hopping term is present
($Y\neq 0$). Notice that, as soon as $X\neq t$, only $| \eta
\rangle_{\pi}$ states could become ground states, the other choices of
$\phi$ in (\ref{ETA}) giving states that cannot be eigenstates of
(\ref{HEH}).

On the contrary, no exact result is available concerning the existence
of the more structured filling dependent superconducting region when
the constraints on the parameters which allow the 1-$d$ integrability are
removed. It is one purpose of the present paper to investigate within
the Hartree-Fock scheme such possibility, as well as to test how the
superconducting region modifies at $T\neq 0$.

Given the relevance of the pair-hopping term to stabilize the
$\eta_\pi$-pairs phase, and in order to make the physical mechanism more
clear, we shall focus on a subcase of the extended model in which, apart
from the pure Hubbard terms, only the pair hopping amplitude is taken
different from zero. This is known in the literature as
Penson-Kolb-Hubbard model. We want to emphasize here that the presence
of other terms in (\ref{HEH}) is not expected to affect our results
in a qualitative way, as other recent numerical studies confirm
\cite{SZA,BEC}. 

In Section \ref{sec-2} we give the Hamiltonian and derive within the
Hartree-Fock scheme the temperature dependent equations for the filling
and the self-consistent superconducting order parameter. In Section
\ref{sec-3} we solve the equations in dimension $2$ and $3$, and show
the temperature and filling dependence of the superconducting phase in
these cases. In Section \ref{sec-4} we add to the Hamiltonian a
next-nearest neighbors contribution to the hopping term, and we show
how this affects the filling and temperature dependence of the
superconducting phase. Finally, in Section \ref{sec-5} we discuss our
results and give some conclusions.

\end{section}
%%%%%%%%%%%%%%%%%%%%%%%%%%%%%%%%%%%%%%%%%%%%%%%%%%%
%%%%%             S E C T I O N   2           %%%%%
%%%%%%%%%%%%%%%%%%%%%%%%%%%%%%%%%%%%%%%%%%%%%%%%%%%
\begin{section}{\boldmath Penson-Kolb-Hubbard model and Hartree-Fock
(HF) approach to the {\large $\pi$}-phase }

The Penson-Kolb-Hubbard Hamiltonian reads
\begin{equation}
H_{PKH} \, = \, -t \sum_{<{\bf i},{\bf j}> } 
\sum_\sigma c^{\dagger}_{{\bf i},\sigma}
c^{}_{{\bf j}, \sigma} +
U \sum_{\bf i} n_{{\bf i},\uparrow}n_{{\bf
i},\downarrow} \, + \, 
Y \, \sum_{<{\bf i},{\bf j}> }
 c^{\dagger}_{{\bf i},\uparrow} 
c^{\dagger}_{{\bf i},\downarrow}  c^{}_{{\bf
j},\downarrow} c^{}_{{\bf j}, \uparrow} \, + \,
\mu \sum_{\bf i} n_{{\bf i}} \quad , 
\label{PKH-Ham}
\end{equation}
where the last term accounts for the chemical potential.
\\The case $U=0$ in 1-$d$ was first examined by Penson and Kolb
\cite{PK} to study a short range interaction between electron pairs of
small radius (actually zero), in contrast with the 
BCS-theory, where the size of the Cooper pairs is comparatively 
large. This led to envisage a {\it real space} formulation 
for the electron pairing, which is very interesting in many contexts of
condensed matter physics.
\\Later on, the Coulomb repulsion term $U$ was also taken into account
by \cite{DON}, where the PKH Hamiltonian was proposed as an effective 
phenomenological model capturing the main physical features 
of doped materials, such as 
high $T_c$ superconductors. Indeed if we assume that, due to some 
(yet unknown) microscopic mechanism, localized pairs can be formed, 
then their displacement in the lattice should be described by a 
pair hopping term competiting with a single carrier hopping 
amplitude $t$. The Coulomb repulsion term should account for the 
insulating phase. 
\\More recently a slave boson wide study of the different possible
phases of the model at zero temperature has also been done 
\cite{POL-1}. In particular, a region characterized by
a non-vanishing value of the order parameter $x_{\pi} = \frac{1}{L^d} \,
< K^{\dagger}_{\pi} >$ was found; here $K^{\dagger}_{\pi}$ is given
by~(\ref{Kphi}) with $\phi=\pi$, and $< \; >$ stands 
for the average value on the grand canonical statistical ensemble. 
In the following we shall denote such a phase as $\pi$-phase.
\\Noticing that $H_{Y}$ in (\ref{HEH}) can be rewritten also as $
-Y \, \sum_{<{\bf i},{\bf j}> }
e^{i \mbox{\small \boldmath $\pi$} \cdot ({\bf i-j})} 
c^{\dagger}_{{\bf i},\uparrow} 
c^{\dagger}_{{\bf i},\downarrow} \,  c^{}_{{\bf
j},\downarrow} c^{}_{{\bf j}, \uparrow}$, 
and assuming a translational invariance throughout the lattice, 
a Hartree-Fock linearization can be implemented on both interaction
terms in (\ref{PKH-Ham}) to study the $\pi$-phase. 
We shall suppose that such a phase is the energetically 
favourite one; therefore we shall take $Y>0$ and assume 
that the thermal energy is not sufficient 
to let other phases emerge (see \cite{POL-1} for a detailed 
study of the possible several phases).
Within this HF approximation the 
PKH Hamiltonian decomposes into a sum of $k$-space independent 
hamiltonians,  
\begin{equation}
H_{PKH} \stackrel{HF}{\approx} \,  
\sum_{{\bf{k}} \in B} H_{\bf{k}} 
\quad , \quad 
[ H_{\bf{k}}, H_{\bf{k^{\prime}}} ]=0 \quad ,
\label{HHF}
\end{equation}
where
\begin{equation}
\normalsize
H_{\bf{k}}= - \left[ (t \epsilon_{\bf{k}}+\mu) 
n_{\bf{k} \uparrow} + 
(t \epsilon_{\small \mbox{\boldmath $\, \pi$} - {\bf k}}+\mu) 
n_{\small \mbox{\boldmath $\, \pi$} - {\bf k} \, \downarrow} 
\right] + 
\tilde{U} \left( x_{\pi} c_{\bf k \uparrow}^{\dagger} 
c_{\small \mbox{\boldmath $\, \pi$} - {\bf k} \, \downarrow}^{\dagger} 
+ x_{\pi}^{*} 
c_{\small \mbox{\boldmath $\, \pi$} - {\bf k} \, \downarrow} 
c_{\bf k \uparrow}\right)  - \tilde{U} |x_{\pi}|^2
\label{Hk}
\end{equation}
and $\tilde{U} = U-qY$, $q$ being the number of nearest 
neighbors, equal to $2 d$ for a hypercubic lattice. 
The sum in (\ref{HHF}) runs over the Brillouin zone 
$B$ and the ${\bf{k}}$ vectors are measured in units 
of the inverse lattice spacing ({\it i.e.} 
$-\pi \le k_i \le \pi$). 
\\In contrast with (\ref{PKH-Ham}), the linearized hamiltonian
(\ref{Hk}) does not preserve the number of particles; indeed in a HF
picture the $\pi$-phase has to be thought of as a superposition of
$\eta_{\pi}$-pairs involving different number of pairs, the
{\it average} number of electrons being fixed through the chemical
potential $\mu$. 
\\Since the dynamical algebra of each $H_{\bf k}$ is the 
$SU (2) \oplus U(1)$ given by the following generators:
\begin{eqnarray}
& J_{\bf{k}}^{\dagger} = c_{\bf k \uparrow}^{\dagger} 
c_{\small \mbox{\boldmath $\, \pi$} - {\bf k} \, 
\downarrow}^{\dagger} \hspace{3cm} \nonumber \\
& J_{\bf{k}}^{-} = 
c_{\small \mbox{\boldmath $\, \pi$} - {\bf k} \, \downarrow} 
c_{\, \bf k \uparrow} 
\hspace{3cm} \nonumber \\
& J_{\bf{k}}^{z} = (n_{\small \mbox{\boldmath $\, \pi$} - 
{\bf k} \, \uparrow} +  
n_{\small \mbox{\boldmath $\, \pi$} - {\bf k} \, \downarrow} -1) 
/ 2 \hspace{.5cm} \nonumber \\
& S_{\bf{k}}^{z} = (n_{\small \mbox{\boldmath $\, \pi$} - 
{\bf k} \, \uparrow} -  
n_{\small \mbox{\boldmath $\, \pi$} - {\bf k} \, \downarrow} ) 
/ 2 \hspace{0.9cm} ,\nonumber 
\end{eqnarray}
we can diagonalize it by means of the following unitary 
transformation, 
\[
H_{\bf{k}} \rightarrow H_{\bf{k}}^{\prime} = 
e^{L_{\bf{k}}} H_{\bf{k}} e^{-L_{\bf{k}}} \hspace{1cm} \quad , \quad  
L_{\bf{k}} = a_{\bf{k}} (x_{\pi} 
J_{\bf{k}}^{\dagger} 
- x_{\pi}^{*} J_{\bf{k}}^{-}) \quad , 
\]
where $a_{\bf{k}}$ are real numbers satisfying the relation 
\[
\tan{(2 a_{\bf{k}} |x_{\pi}| )} = 
- \frac{|x_{\pi}| \tilde{U}}{\mu+ t 
(\epsilon_{\bf{k}}+ 
\epsilon_{\small \mbox{\boldmath $\, \pi$} - {\bf k}}) /2} \quad .
\]
The grand partition function $Z$ in the HF approximation can then be
worked out, since  
\[ Z = Tr \, e^{- \beta H_{PKH}} \, \stackrel{HF}{\approx}  
\prod_{\bf{k}} Tr_{\bf{k}} e^{-\beta H_{\bf{k}}} = 
\prod_{\bf{k}} Tr_{\bf{k}} e^{-\beta H_{\bf{k}}^{\prime}} \quad , \]
where $Tr _{\bf{k}}$ is the trace over the space of quantum 
labels $(\bf{k}, \uparrow)$ and 
$(\mbox{\boldmath $\pi$} - {\bf k} , \downarrow)$. 
\\The thermodynamic grand potential (per particle) 
in the thermodynamic limit now reads 
\begin{eqnarray}
\lefteqn{\omega = \lim_{T.D.} - k_{B}T \, \frac{\ln Z}{N} \, =
\hspace{10cm} \vspace{1cm} }   \\ & & 
= 
- \frac{1}{(2 \pi)^d} \int_{B} d {\bf k} \left[ A_{\bf{k}} +  
k_{B} T \, \ln [2 \cosh  \beta \frac{D_{\bf{k}} + 
R_{\bf{k}}}{2} ]
+ \ln [2 \cosh  \beta \frac{D_{\bf{k}}- R_{\bf{k}}}{2} ]
\frac{}{} \right] \; , \nonumber
\end{eqnarray}
with
\begin{eqnarray}
& S_{\bf{k}} = t(\epsilon_{\bf{k}}+ 
\epsilon_{\small \mbox{\boldmath $\, \pi$} - {\bf k} }) /2 
\nonumber \\
& D_{\bf{k}} = t(\epsilon_{\bf{k}}- 
\epsilon_{\mbox{\small \boldmath $\, \pi$} - {\bf k}}) /2 
\label{SDAR} \\
& A_{\bf{k}} = \mu + S_{\bf{k}} + 
\tilde{U} |x_{\pi}|^{2}  \nonumber \\
& \hspace*{1.0cm} R_{\bf{k}} = \sqrt{(\mu + S_{\bf{k}})^2 + 
\tilde{U}^{2} |x_{\pi}|^{2}} \quad .
\nonumber 
\end{eqnarray}
In order to investigate the thermodynamical properties 
of the system one has to implement the selfconsistency equation 
${\partial \omega}/{\partial x_{\pi}} = 0$ 
for the order parameter, which turns out to be 
\begin{displaymath}
x_{\pi} \, \cdot \, \int_{B} d {\bf k} 
\left[ 
1+ \frac{\tilde{U}}{4 R_{\bf{k}}} 
[\tanh  \beta \frac{D_{\bf{k}}+ R_{\bf{k}}}{2} - 
\tanh  \beta \frac{D_{\bf{k}}- R_{\bf{k}}}{2} ] \right] = 0 \quad .
\end{displaymath}
As usual, this equation gives a solution 
$x_{\pi} \equiv 0$ for $T \ge T_c$ and a solution $x_{\pi} \neq 0$ 
for $T \le T_c$. It is easy to see that the $\pi$-phase 
exists only when $\tilde{U}=U-qY \le 0$, 
that is when the pair-hopping $Y>0$ term renormalizes the 
effective interaction $U>0$ to an {\it attractive} regime. 
Indeed for $T \le T_c$
the above equation gives:
\begin{equation}
\tilde{U}^{-1} = -  
\frac{1}{(2 \pi)^d} \int_{B} d {\bf k} \,
\frac{1}{4 R_{\bf{k}}} \left[ 
[\tanh  \beta \frac{D_{\bf{k}}+ R_{\bf{k}}}{2} - 
\tanh  \beta \frac{D_{\bf{k}}- R_{\bf{k}}}{2} ] 
\right] = 0 \quad .
\label{autoc-eq}
\end{equation}
Moreover, one must also satisfy the filling equation, 
$n = {\partial \omega}/{\partial \mu}$, 
which reads
\begin{equation}
n=1+ \frac{1}{(2 \pi)^d} \int_{B} d {\bf k} \, 
\frac{\mu + S_{\bf{k}}}{2 R_{\bf{k}}} 
[\tanh  \beta \frac{D_{\bf{k}}+ R_{\bf{k}}}{2} - 
\tanh  \beta \frac{D_{\bf{k}}- R_{\bf{k}}}{2} ] \quad .
\label{fill-eq}
\end{equation}
\\Eqns. (\ref{autoc-eq})-(\ref{fill-eq}) constitute the 
{\it parametric} form of the equation of state. 
In order to get to one {\it closed} form, one should in 
principle invert (\ref{fill-eq}) obtaining $\mu$ as a function 
of $n$, $T$ and $\tilde{U}$, and then insert it 
into~(\ref{autoc-eq}). The thermodynamics of the model 
will then be expressed in terms of $n, T$ and $\tilde{U}$.
\\Noticeably, comparing (\ref{autoc-eq}) and (\ref{fill-eq}) 
it is easy to show that whenever $S_{\bf{k}}$ in 
(\ref{SDAR}) vanishes, {\it i.e.} whenever:
\begin{equation}
\epsilon_{\small \mbox{\boldmath $\, \pi$} - {\bf k} } = 
- \epsilon_{\bf{k}}
\label{S0}
\end{equation}
the chemical potential can be {\it exactly} inverted. 
In this case we simply have:
\begin{equation}
\mu= \frac{1-n}{2} \, \tilde{U}
\label{mu}
\end{equation}
We wish to stress here that for a given model, even within 
the HF approximation, it is not obvious at all that $\mu$ can be
inverted exactly: in the PKH model it is a peculiar feature of 
the $\pi$-phase (which is not shared by $| \eta \rangle_{0}$ phase). 
\\The relations (\ref{S0})-(\ref{mu}) hold in any dimension for a
hypercubic lattice when dealing with a {\it nearest neighbors}
hopping term, since the dispersion relations reads:
\begin{equation}
\epsilon_{\bf{k}} = \sum_{i=1}^{d} 2 \, \cos{k_i}
\label{eps-nn}
\end{equation}
As we will discuss in \ref{sec-4}, eqn.(\ref{S0}) 
is no more satisfied when a {\it next nearest neighbors} 
term is included.
\label{sec-2}
\end{section}
%%%%%%%%%%%%%%%%%%%%%%%%%%%%%%%%%%%%%%%%%%%%%%%%%%%
%%%%%             S E C T I O N   3           %%%%%
%%%%%%%%%%%%%%%%%%%%%%%%%%%%%%%%%%%%%%%%%%%%%%%%%%%
\begin{section}{\boldmath The temperature and filling 
dependence of the phase diagram in $d=2$ and $d=3$}
\noindent In this section we consider the case of a nearest neighbors 
hopping term, the dispersion relation being given by (\ref{eps-nn}). 
We aim to derive the features of 
the critical values of $\tilde{U}$ versus $n$ for a given 
temperature $T$, in order to compare them with known solution 
of similar models (see section \ref{sec-5}). 
Such a critical curve $\tilde{U}_c = \tilde{U}_c (n)$ is obtained 
from (\ref{autoc-eq})-(\ref{fill-eq}) by setting the 
order parameter $x_{\pi}=0$ into (\ref{SDAR}).
\\As already noticed in section \ref{sec-2}, besides the parametric 
form (\ref{autoc-eq})-(\ref{fill-eq}), in this case we can also deal 
with one closed form; indeed, since (\ref{S0}) is satisfied, 
$R_{\bf{k}}$ is independent of ${\bf{k}}$, and therefore (\ref{mu}) 
holds. By substituting $\mu$ 
into $R_{\bf{k}}$, and then $R_{\bf{k}}$ into (\ref{autoc-eq}), 
we obtain from (\ref{fill-eq}) the critical equation  
\begin{equation}
\frac{1}{(2 \pi)^d} 
\int_{B} d {\bf{k}} \, \frac{1}{2} \left[ \tanh{[\beta  \, 
\frac{t \epsilon_{\bf{k}} + \tilde{U}_{c} \, \delta(n)/2}{2} ]} -
\tanh{[\beta  \, \frac{t \epsilon_{\bf{k}} - 
\tilde{U}_{c} \, \delta(n)/2}{2} ]} \quad ,  
\right] = \delta(n)
\label{cl-1}
\end{equation}
where $\delta(n)=1-n$ is the `doping' and $B$ is the 
$d$-dimensional Brillouin zone.
\\In order to work out the integration it is worth to 
introduce the `density of states' 
$g^{(d)}(\epsilon)$ with normalization  
\begin{equation}
\frac{1}{(2 \pi)^d} \int_{B} d {\bf{k}} = 
\int_{\epsilon_{min}}^{\epsilon_{max}} 
d\epsilon \, g^{(d)}(\epsilon) \quad ,
\label{normal-g}
\end{equation}
which turns out to be in this case an 
even function of $\epsilon$, so that 
$\epsilon_{min}=-\epsilon_{max}$ (this will not be 
the case in section \ref{sec-4}).
Moreover we have $\epsilon_{max}= 2 d$.
\\In $d=2$ it is known \cite{POL-2} that 
\begin{equation}
g^{(2)}(\epsilon) = \frac{1}{2 \pi^2} 
K \left(1-\left(\frac{\epsilon}{4} \right)^2 \right) \quad ,
\label{g2}
\end{equation}
where $K$ is the complete elliptic integral of the 
first kind; this function has a logarithmic divergence 
at $\epsilon=0$.
\\For the three-dimensional case we have also performed 
the calculation, obtaining 
%\[
%g^{(3)}(\epsilon)= \left\{ 
%\begin{array}{ll} 
%\frac{1}{\pi^3} \int_{0}^{\arccos(\epsilon/2-2)}  
%K(1- (\frac{\epsilon/2-\cos(x)}{2})^2) dx
%& \mbox{for: } 2 \le | \epsilon| \le 6 \\ 
%\frac{1}{\pi^3} \int_{0}^{\pi} 
%K(1- (\frac{\epsilon/2-\cos(x)}{2})^2) dx & 
%\mbox{for: } 0 \le | \epsilon| \le 2 
%\end{array} \right.
%\]
%%%%%%%%%%%%%%%%%%%%%%%%%%%%%%%%%%%%%%%%%%%%%%%%%%%%%%%%%%%%%
\begin{eqnarray}
 & {\displaystyle \frac{2}{\pi} \int_{0}^{\pi} 
g^{(2)}(\epsilon /2- \cos(x)) dx} \hspace{1.5cm}  & 
\hspace{.5cm} 0 \le | \epsilon| \le 2 \nonumber \\
g^{(3)}(\epsilon): =  \hspace{.5cm} & \hspace{3cm}  & \hspace{2cm}  \\
& {\displaystyle \frac{2}{\pi} \int_{0}^{\arccos(\frac{\epsilon}{2}-2)}  
g^{(2)}(\epsilon /2- \cos(x)) dx}
& \hspace{.5cm} 2 \le | \epsilon| \le 6 \quad , \nonumber
\end{eqnarray}
%%%%%%%%%%%%%%%%%%%%%%%%%%%%%%%%%%%%%%%%%%%%%%%%%%%%%%%%%%%%%
which is plotted in fig.\ref{pkh_fig2}.
\\Since eqn.(\ref{cl-1}) is invariant under the transformation 
$\delta(n) \rightarrow - \delta(n)$, it is easily seen that 
$\tilde{U}_c$ is symmetric with respect to the value at half
filling ($n=1$). The critical curves in $d=2$ 
and $d=3$ are plotted in figs.\ref{pkh_fig3} and \ref{pkh_fig4} 
respectively. 
For the sake of consistency with our approximation, 
we have plotted the region of the phase 
diagram where the values of $\tilde{U}$ do 
not exceed the band-width $\Delta= 4 t \, d$. 
\\Notice the different behaviour of $\tilde{U}_c$ in the 
two cases, in particular for low temperatures. Indeed in 
$d=2$ we have a very sharp, cuspid-like shape at half 
filling, while in $d=3$ a `plateau' is obtained, meaning 
that the effective interaction threshold is almost independent 
of the density of electrons in the lattice for a rather 
wide range of $n$. 
This effect is due to the quite different behaviour of 
the density of states $g^{(2)}$ and $g^{(3)}$. Indeed 
using the parametric form it is possible to show that 
at low $T$'s the shape of $\tilde{U}_c$ in the neighborhood 
of half filling is governed by the behaviour of $g^{(d)}$ 
around $\epsilon=0$. In fact from eqns. (\ref{mu}) and (\ref{cl-1})
we easily obtain that the value of $\mu$ is determined by the
equation
\begin{equation}
\int_{-\epsilon_{max}}^{+\epsilon_{max}} g^{(d)}(\epsilon) \, 
P_{\beta}(\epsilon; \mu) = 1-n \quad , 
\label{Uexact}
\end{equation}
where $\beta=1/{k_{B} T}$, and $P_{\beta}(\epsilon; \mu)$ is 
\[
P_{\beta}(\epsilon; \mu)=\frac{1}{2} 
[\tanh  \beta t \frac{\epsilon -\mu}{2} - 
\tanh  \beta t \frac{\epsilon+\mu}{2} ] \quad .
\]
Taking into account the normalization (\ref{normal-g}) of 
$g^{(d)}(\epsilon)$ and the fact that $P_{\beta}(\epsilon;\mu)$ 
is positive and not greater than $1$, one can easily see that 
$n=1$ implies $\mu=0$ (for any temperature $T$). Since from
eqn.(\ref{mu}) we have $\tilde{U}=2 \mu / (1-n)$, 
a simple limit $\mu \rightarrow 0$ can be performed, 
yielding
\begin{equation}
\tilde{U}_c (n=1) \, = \, - \left( 
{\int_{-\epsilon_{max}}^{+\epsilon_{max}} g^{(d)}(\epsilon)} 
\, p_{\beta}(\epsilon) \right)^{-1} \label{Uc-n1} \quad , 
\end{equation}
where
\begin{equation}
p_{\beta}(\epsilon) = \frac{\beta t}{4} \cosh^{-2} 
(\frac{\beta t}{2} \, \epsilon) \label{p_beta} \quad .
\end{equation}
This holds for {\it any} temperature $T$. In particular when 
$T \rightarrow 0$ one gets 
\begin{equation}
\tilde{U}_c (n=1) \, = \, - \, \frac{1}{g^{(d)}(0)}
\label{Uc-n1-T0}
\end{equation}
Now, since $g^{(3)}$ is almost 
constant around $\epsilon =0$, when $T \sim 0$ eqn. 
(\ref{Uexact}) gives
\begin{equation}
1-n \, \approx \, 
\displaystyle \int_{-|\mu|}^{+|\mu|} g^{(3)}(\epsilon) \, 
\, \approx \, - 2 \mu \, \left. g^{(3)}(\epsilon) 
\right|_{\epsilon =0}
\quad , \end{equation}
and using (\ref{mu}) we obtain that $\tilde{U}_c$ is actually 
independent of $n$ (indeed $g^{(3)}(0) = 0.1427$ and so 
$1/0.1427 = 7.0078$ which is just the value of $\tilde{U}$ 
around half filling when $T=0$).
To be more precise, since the value of $g^{(3)}$ at 
$|\epsilon|=2$ is slightly higher than at $\epsilon=0$ 
(see fig.\ref{pkh_fig2}), 
we have that at $T=0$ the plateau is slightly 
concave, so that the highest value of $\tilde{U}_c$ is 
actually reached away from half filling, at the symmetric 
values $n \sim 0.4$ and $n \sim 1.6$.
\\Eqn.(\ref{Uc-n1-T0}) stems from eqn.(\ref{Uc-n1}) because the 
functions $p_{\beta}(\epsilon)$ are `regularizing' 
in the sense that they are positive and their integral on 
$\mathcal R$ equals $1$ for any $\beta$ (though strictly 
speaking they belong to $\mathcal{S}(R)$ and not to 
${\mathcal{C}}^{\infty}_{0}({\mathcal R})$). 
In fact in $d=2$ we observe that, when 
$\beta \rightarrow \infty$, 
\begin{eqnarray}
\int_{-4}^{4} g^{(2)}(\epsilon) \, p_{\beta}(\epsilon) 
\, \ge \, 
\int_{-2/(\beta t)}^{2/(\beta t)} g^{(2)}(\epsilon) \, 
p_{\beta}(\epsilon) \, \ge \hspace{5cm}  \nonumber \\
\ge \,
\underbrace{\int_{-2/(\beta t)}^{2/(\beta t)} 
p_{\beta}(\epsilon)}_{2 \tanh{1}}  \, \cdot \,  
\inf_{0 \le |\epsilon| \le 2/(\beta t)} g^{(2)} \, \, 
\stackrel{\beta \rightarrow \infty}{\longrightarrow} 
\, \infty \quad ,
\end{eqnarray}
because of the divergence of $g^{(2)}$ at $\epsilon=0$.
\\The 3-$d$ case is slightly different because $g^{(3)}$ 
is finite at $\epsilon=0$; nevertheless, if we take a 
sufficiently small $\delta$ around $\epsilon=0$, it easily 
realized that 
\[
\lim_{\beta \rightarrow \infty} \int_{\delta \le |\epsilon| 
\le \epsilon_{max}} g^{(3)}(\epsilon) \, p_{\beta}(\epsilon) 
d \epsilon \, = \, 0 \, = \, \lim_{\beta \rightarrow \infty} 
\int_{\delta \le |\epsilon| \le \epsilon_{max}}  \, 
p_{\beta}(\epsilon) d \epsilon \quad , \] 
which yields
\[
\inf_{|\epsilon| \le \delta} g^{(3)} \, \le \, 
\lim_{\beta \rightarrow \infty} 
\int_{- \epsilon_{max}}^{+ \epsilon_{max}} g^{(3)}(\epsilon) 
\, \, p_{\beta}(\epsilon) \, \,  \le \, \, 
\sup_{|\epsilon| \le \delta} g^{(3)} \hspace{2cm} \forall 
\delta>0 \quad , \]
and so again (\ref{Uc-n1-T0}).
\label{sec-3}
\end{section}
%%%%%%%%%%%%%%%%%%%%%%%%%%%%%%%%%%%%%%%%%%%%%%%%%%%
%%%%%             S E C T I O N   4           %%%%%
%%%%%%%%%%%%%%%%%%%%%%%%%%%%%%%%%%%%%%%%%%%%%%%%%%%
\begin{section}{The next nearest neighbors contribution}
\noindent Let us now turn to the case when in the Hamiltonian
(\ref{PKH-Ham}) a next 
nearest neighbors contribution is included in the hopping term,
which therefore becomes
\begin{equation}
 -t \sum_{<{\bf i},{\bf j}> } 
\sum_\sigma c^{\dagger}_{{\bf i},\sigma}
c^{}_{{\bf j}, \sigma} -\alpha \, t \sum_{<<{\bf i},{\bf j}>> } 
\sum_\sigma c^{\dagger}_{{\bf i},\sigma}
c^{}_{{\bf j}, \sigma} \quad ,
\end{equation}
Notice that such a term explicitly breaks the particle-hole 
$c^{\dagger}_{\bf j} \rightarrow e^{i \mbox{\small \boldmath $\pi$}
\cdot {\bf j}}  c^{}_{\bf j} $ invariance of the model. 
The dispersion relation is now given by 
\begin{equation}
\epsilon_{\bf{k}} = \sum_{i=1}^{d} 2 \, \cos{k_i} + 
\alpha \prod_{i < j \le d} 4 \, \cos{k_i} \cos{k_j} \quad .
\label{eps-nnn}
\end{equation} 
As observed at the end of section \ref{sec-2}, the symmetry 
$\epsilon_{\bf{k}} =- \epsilon_{\mbox{\small \boldmath 
$\pi$}-{\bf{k}}}$ does not hold anymore; as a consequence, 
$S_{\bf k}$ in eqn.(\ref{SDAR}) is no 
more vanishing, and $R_{\bf k}$ does 
depend on $\bf k$. This yields both mathematical and 
physical new features. In particular, 
the equation for the critical surface has to be given 
in the parametric form (\ref{autoc-eq})-(\ref{fill-eq}), 
$\mu$ being the parameter.
\\We are here interested in the case $d=2$. 
We have computed also in this case the density of 
states, that turns out to be 
\begin{equation}
g^{(2)}_{\alpha}(\epsilon) = 
\frac{1}
{2 \pi^{2} \sqrt{1+ \alpha \epsilon}} \,
K \left( \frac{1-( \epsilon/4- \alpha)^2}
{1+ \alpha \epsilon} \right) \quad ,
\end{equation}
where $\alpha$ is assumed to be $|\alpha| \le 1/2$, since 
the next-nearest neighbors term is expected to be small 
(in $L^2$) with respect to 
the nearest neighbors one. For $\alpha=0$ 
we recover the usual form (\ref{g2}). 
Notice also that the function $g^{(2)}_{\alpha}$ 
is no more even in $\epsilon$; indeed 
we now have $g^{(2)}_{\alpha}(-\epsilon)=g^{(2)}_{-\alpha}
(\epsilon)$. The logarithmic 
divergence is shifted to $\epsilon=-4 \alpha$, and the 
domain is characterized by $\epsilon_{min}=-4(1- \alpha)$ 
and $\epsilon_{max}=4 (1+ \alpha)$, as shown in fig.\ref{pkh_fig5}.
\\This imparts the critical curve $\tilde{U}_c$ vs $n$ an 
asymmetric form, the highest value of $\tilde{U}$ falling 
now at a $n_{max} \neq 1$, as shown in fig.\ref{pkh_fig6}. 
Moreover, at a given temperature, such a maximum of the critical 
curve is shifted upward with respect to the curve of the case 
$\alpha=0$. This means that at a given temperature $T$, the 
effect of the next nearest neighbors term is to {\it reduce} 
the `optimal' effective attraction $\tilde{U}$. In turn, this implies
at a given $\tilde{U}$ the {\it raising} of the highest critical
temperature reachable by doping the system.  
\\We have also studied 
how $n_{max}$ depends on $\alpha$. The relation is almost 
linear for $-0.4 \le \alpha \le 0.4$, while it displays a sudden 
increase around $\alpha \sim 0.45$; in fig.\ref{pkh_fig7} we 
have extended the curve to the range $0 \le |\alpha| \le 1$ 
(which could be still acceptable in principle) to show how $n_{max}$ 
saturates towards the limiting values $n=0$ or $n=2$. Notice 
that the curve is odd in $\alpha$; this is because the 
parametric eqns.(\ref{autoc-eq})-(\ref{fill-eq}) are invariant under 
the transformation 
$\mu \rightarrow -\mu \, ; \, \alpha \rightarrow -\alpha \, ; 
\,\tilde{U} \rightarrow \tilde{U}$, and therefore one can 
show, in agreement with \cite{POL-2}, that
\begin{displaymath}
T_c \, (\tilde{U},n;\alpha) = T_c \,(\tilde{U},2-n;-\alpha) \quad .
\end{displaymath}
It is possible to see that the curve 
depends very weakly on the temperature $T$. 
\\Finally, we have plotted the phase diagram of $T_{c}$ versus 
$n$ at fixed $\alpha$, which is shown in fig.\ref{pkh_fig8}.
We can observe that the next nearest neighbors term has 
mainly two effects. First it shifts away from half filling 
the range of values of $n$ at which the superconducting phase 
exists; this suggests that the system has to be doped in order to 
observe a superconducting behaviour. Secondly, it raises up the 
highest reachable critical temperature with respect to the case 
where only a nearest neighbors interaction is considered.
Having in mind the phenomenology of high $T_c$ materials, this
study supports the idea that the actual microscopic 
Hamiltonian should be particle-hole {\it not}-invariant.  
\label{sec-4}
\end{section}
%%%%%%%%%%%%%%%%%%%%%%%%%%%%%%%%%%%%%%%%%%%%%%%%%%%
%%%%%             S E C T I O N   5           %%%%%
%%%%%%%%%%%%%%%%%%%%%%%%%%%%%%%%%%%%%%%%%%%%%%%%%%%
\begin{section}{Discussion and conclusions}
The phase diagrams obtained within the Hartree-Fock scheme can be given
a more precise physical interpretation by comparing them with the $1$-d
exact results known for some very specific cases.
Strictly speaking, such a comparison is only possible for $T=0$, since
the phase diagram of these integrable models is not known at $T \neq 0$.
In so doing, we observe that at zero temperature our results in
fig.\ref{pkh_fig3} and fig.\ref{pkh_fig4} have the same structure as
those in fig.\ref{pkh_fig1}. One can recognize three distinct regions
in the phase space. The first one is characterized by filling
independent $\pi$-phase.
By reducing the effective attraction $\tilde U$, one enters a second
region in which the existence of the $\pi$-phase depends on the actual
filling. Finally, above the critical curve the $\pi$-phase disappears.
\\In analogy with what is known to happen for the solved $1$-d cases,
we can think of the filling independent region as the phase in which all
particles are paired in $\eta_\pi$-pairs (\ref{ETA}). This is
in agreement with known result in $d>1$ \cite{MON}. The second region
should be characterized by simultaneous presence of paired and
unpaired electrons and empty sites, whereas in the third one no paired
electrons could move.  

Switching on the temperature, thermal fluctuations are expected to
break pairs. The dependence on the temperature of our phase diagrams
supports this idea. Actually for a given filling $n$, the greater
becomes the temperature $T$, the greater must be the magnitude of the
effective attractive interaction $\tilde{U}$ in order to keep the
$\eta_\pi$-pairs bound together.
In fact this is what our phase diagrams in figs. \ref{pkh_fig3} 
and \ref{pkh_fig4} show: the 
curves of higher $T$'s lay below the lower $T$'s ones; this can 
also be demonstrated analitically very easily, at least in the 
cases of a nearest neighbors hopping term. Indeed from eqn.(\ref{cl-1}), 
which is 
of the form $G(n,\beta,\tilde{U}_{c})=1-n$, we obtain that 
\begin{equation}
\frac{d \tilde{U}_{c}}{d T} =  - \frac{1}{k_{B} T^2} 
\frac{d \tilde{U}_{c}}{d \beta} = 
- \frac{1}{k_{B} T^2} \frac{\partial G / \partial \beta}
{\partial G / \partial \tilde{U}_{c}} \quad .
\label{dini}
\end{equation}
By working out the derivatives, and using the fact that the functions 
$\tanh(x)$ and $x/\cosh^{2}(x)$ are monotone and that $\tilde{U}$ 
is negative, it is easily seen that (\ref{dini}) is always~$\le 0$.
\\It is important to recall that, thanks to the presence of the
pair-hopping term, an effective attractive interaction $\tilde U$ is
consistent with a positive value of the Coulomb interaction $U$. Hence
the present HF treatment of the thermodynamics of the PKH model 
yields a structured filling dependent superconducting phase even
in presence of repulsive on site Coulomb interaction between electrons.

Finally, we wish to stress that the phase diagram in fig.\ref{pkh_fig8} 
--obtained by including the next nearest neighbors hopping term-- 
exhibits appealing features: for $\alpha\neq 0$ the optimal doping of
the superconducting region is at $n_{max}\neq 1$, and the critical
temperature is enhanced. Moreover, with respect to the results reported
in \cite{POL-2} on the attractive Hubbard model, our figure shows that
even at $T=0$ the superconducting phase exists only for an appropriate
range of filling values, {\it not} including half-filling. In
fact, one could observe that the whole curve of the critical temperature
vs. filling (\ref{pkh_fig8}) is very reminiscent of that obtained for
high-$T_c$ materials.
 
The study of the influence that particle-hole non symmetric terms in 
the Hamiltonian have on the features of the phase diagram has been
worked out in 2-$d$, the conduction in high-$T_c$ superconducting 
materials tipically taking place along the cuprate planes. As
Hartree-Fock approach is more accurate the higher is the dimension,
dealing with a 3-$d$ and {\it anisotropic} order parameter would 
possibly be more reliable. Work is in progress along these
lines. At the same time, since the results obtained here are
encouraging, a numerical study of the temperature behavior of the
present model in $d=2$ would be probative. 
\label{sec-5}
\end{section}
%%%%%%%%%%%%%%%%%%%%%%%%%%%%%%%%%%%%%%%%
%%%%%      R E F E R E N C E S     %%%%%
%%%%%%%%%%%%%%%%%%%%%%%%%%%%%%%%%%%%%%%%

%%%%%%%%%%%%%%%%%%%%%%%%%%%%%%%%%%%%%%%%%%%%%%%%%
%%%%%    F I G U R E    C A P T I O N S     %%%%%
%%%%%%%%%%%%%%%%%%%%%%%%%%%%%%%%%%%%%%%%%%%%%%%%%
%%%%%%%%%%%%%%%%%%%%%%%%
%%%%%   FIGURE 1   %%%%%
%%%%%%%%%%%%%%%%%%%%%%%%
\begin{figure}
\caption{\noindent Phase diagram in $d=1$ at $T=0$ of 
AAS model (solid curve) and EKS model (dot-dashed curve).
A filling-independent (superconducting) region is obtained
in which the $\eta_{\alpha}$ pairs are ground state. Above it,
a filling-dependent region (still superconducting) exists.
In the EKS model the presence of a {\it pair hopping} term contributes
to extend the superconducting zone towards positive values 
of $U$. In such exactly solved 1-$d$ models 
the phase diagram reaches a maximum around half filling; 
moreover some unphysical constraint (for instance $X=t$) 
have to be imposed on the coupling constants in order to make 
the Hamiltonian integrable.}
\label{pkh_fig1}
\end{figure}
%%%%%%%%%%%%%%%%%%%%%%%%
%%%%%   FIGURE 2   %%%%%
%%%%%%%%%%%%%%%%%%%%%%%%
\begin{figure}
\caption{\noindent The density of states in $d=3$ of the 
hopping term; notice that the value of $g^{(3)}$ in the 
cuspids at $|\epsilon|=2$ is slightly higher than at 
$\epsilon=0$; this yields (see fig.\ref{pkh_fig4}) that 
at $T=0$ the maximum of the critical curve is reached 
away from half filling.}
\label{pkh_fig2}
\end{figure}
%%%%%%%%%%%%%%%%%%%%%%%%
%%%%%   FIGURE 3   %%%%%
%%%%%%%%%%%%%%%%%%%%%%%%
\begin{figure}
\caption{\noindent Phase diagram of the $\pi$-phase in $d=2$; 
the critical value $\tilde{U}_c$ of the effective attraction 
$\tilde{U}=U-qY$ is plotted as a function of the filling for 
some values of the temperature $T$. At a given temperature 
the $\pi$-phase exists for $\tilde{U} \le \tilde{U}_c \,$ ; 
here  $\Delta=8 t$ is the bandwidth and $q=4$ is the number 
of nearest neighbors in the lattice. The values of $\tilde{U}$ 
are negative because the pair hopping term renormalizes the 
on-site Coulomb repulsion $U$ to a negative regime. Notice 
also that the curves are centered around half filling; at 
$T=0$ a filling independent region exists for 
$\tilde{U} \le -\Delta$, like in AAS and EKS models 
(see fig.\ref{pkh_fig1}).}
\label{pkh_fig3}
\end{figure}
%%%%%%%%%%%%%%%%%%%%%%%%
%%%%%   FIGURE 4   %%%%%
%%%%%%%%%%%%%%%%%%%%%%%%
\begin{figure}
\caption{\noindent Phase diagram of the $\pi$-phase in 
$d=3 \,$; $\Delta=12 t$ is the bandwidth and $q=6$ is the 
number of nearest neighbors in the lattice. With respect to 
the bidimensional case (see fig.\ref{pkh_fig3}) the curves 
have a plateau around half filling; indeed at $T=0$ the 
maximum values of $\tilde{U}_c$ are reached at the symmetric 
values $n \sim 0.4$ and $n \sim 1.6$. As the temperature is 
increased, the $\eta_{\pi}$-pairs become more likely to be 
broken and the extension of the $\pi$-phase reduces.}
\label{pkh_fig4}
\end{figure}
%%%%%%%%%%%%%%%%%%%%%%%%
%%%%%   FIGURE 5   %%%%%
%%%%%%%%%%%%%%%%%%%%%%%%
\begin{figure}
\caption{\noindent The density of states in $d=2$ of the 
hopping term when a next nearest neighbors term of coupling 
constant $\alpha$ is also taken into account (solid line); 
notice the asymmetric behaviour with respect to the usual 
case $\alpha=0$: the logarithmic divergence is shifted to 
$\epsilon=-4 \alpha$. This effect is due to the break up 
of the particle-hole invariance of the hopping term; it 
yields an asymmetric shape of the phase diagram 
(see fig.\ref{pkh_fig6}).}
\label{pkh_fig5}
\end{figure}
%%%%%%%%%%%%%%%%%%%%%%%%
%%%%%   FIGURE 6   %%%%%
%%%%%%%%%%%%%%%%%%%%%%%%
\begin{figure}
\caption{\noindent The $\pi$-phase diagram in $d=2$ for different 
values of the next-nearest neighbors hopping amplitude $\alpha$ at 
the temperature $k_{B}T/t=0.1$. 
Notice that when $\alpha$ is increased the 
superconducting region rises up towards less negative values 
of the $\tilde{U}=U-qY$, and its maximum is reached at a $n_{max}$ 
which moves away from half filling. 
This means that, at a given temperature, the next-nearest neighbors 
interaction reduces the effective attraction $\tilde{U}$, yielding 
an increase of the highest reachable critical temperature (see also
fig.\ref{pkh_fig8}). 
}
\label{pkh_fig6}
\end{figure}
%%%%%%%%%%%%%%%%%%%%%%%%
%%%%%   FIGURE 7   %%%%%
%%%%%%%%%%%%%%%%%%%%%%%%
\begin{figure}
\caption{\noindent The behaviour of $n_{max}$ ({\it i.e.} the filling
at which the critical curve $\tilde{U}_c$ vs $n$ in fig.\ref{pkh_fig6}
reaches the maximum) as a function of the next nearest neighbors
coupling constant $\alpha$ (in units of $t$), at temperature 
$k_{B}T/t=0.1$. The curve is odd. Notice that
for $-0.4 \le \alpha \le 0.4$ the behaviour is almost linear; as
$|\alpha|$ is furtherly increased, $n_{max}$ saturates to $0$ or to
$2$, as one can also see in fig.\ref{pkh_fig6}. It is possible to show
that this behaviour of $n_{max}$ depends very weakly on the
temperature~$T$.}
\label{pkh_fig7}
\end{figure}
%%%%%   FIGURE 8   %%%%%
\begin{figure}
\caption{\noindent The critical temperature as a function of the
filling in $d=2$, for a given value $\tilde{U}=-4 t$ of the effective 
attractive interaction. Here $\alpha$ is the coupling constant of the 
{\it next nearest neighbors} hopping term, which breaks the particle-hole 
symmetry
of the model. Notice that, with respect to the case $\alpha=0$, the
next nearest neighbors term yields both the increase of the highest
critical temperature and the displacement away from half filling of
the superconducting $\pi$-phase. In fact the highest~$T_c$ is reached
at $n \sim 1.3$.}
\label{pkh_fig8}
\end{figure}
%%%%%%%%%%%%%%%%%%%%%%%%%%%%%%%%%%%%%%%%%%%%%%
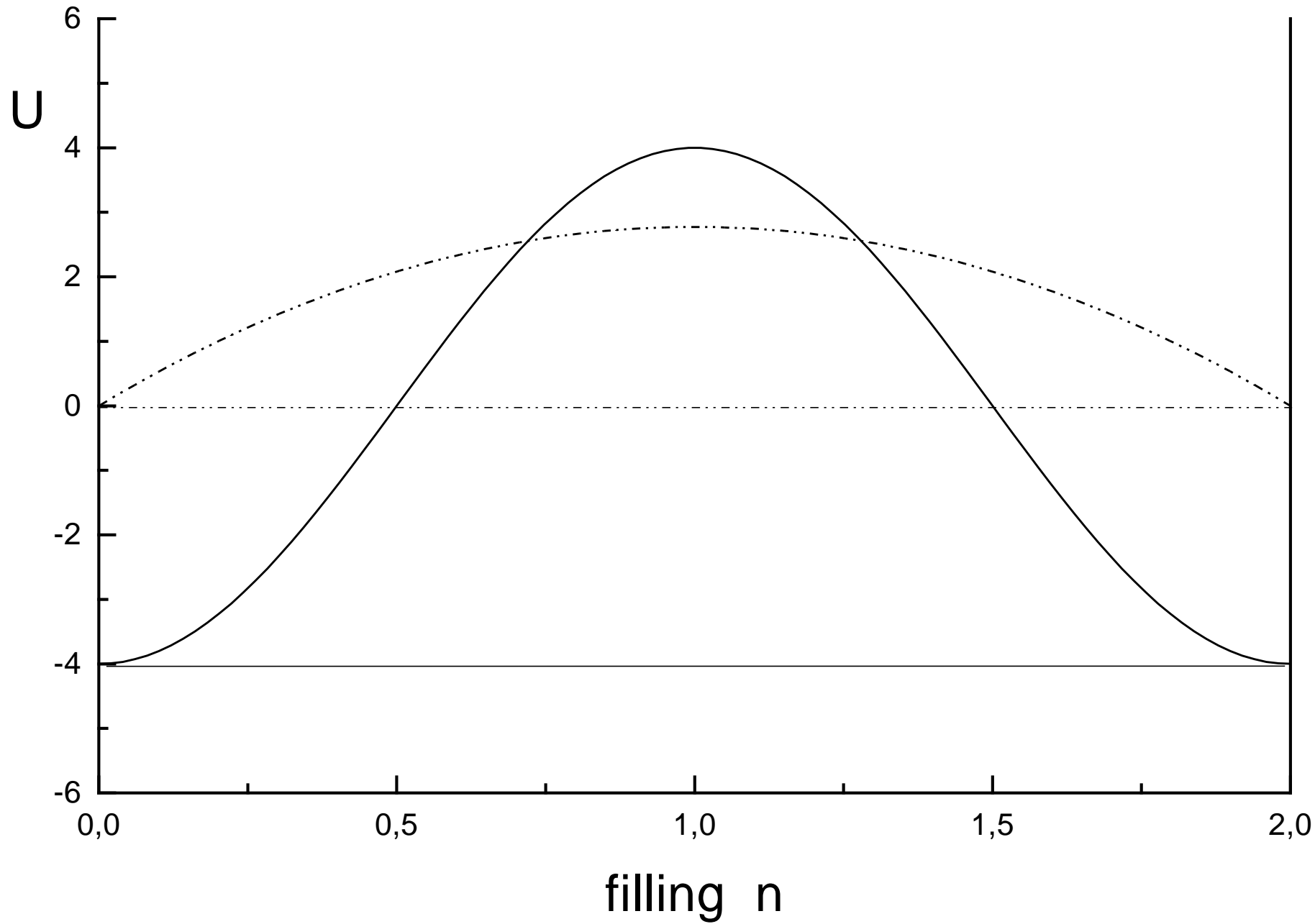
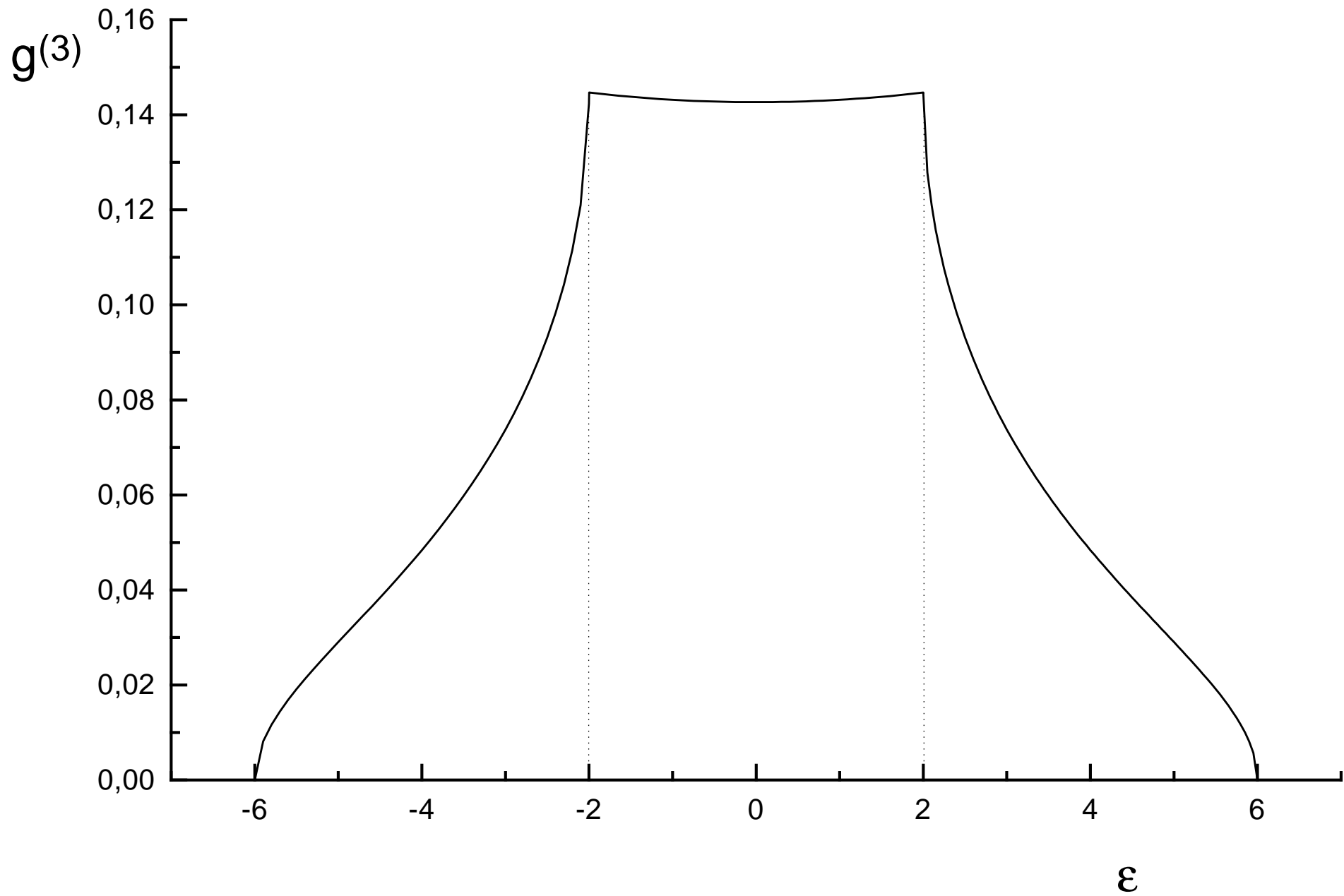
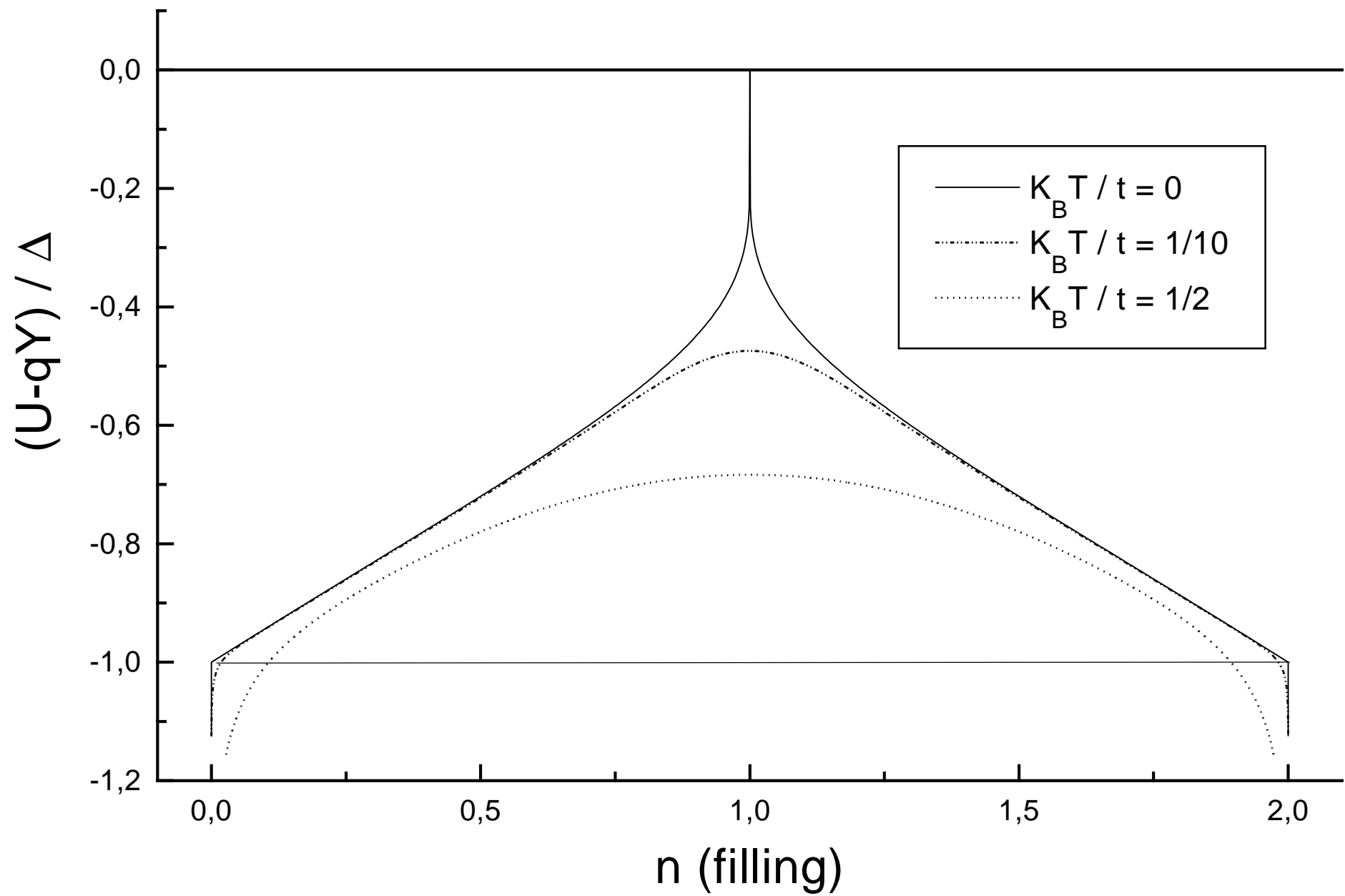
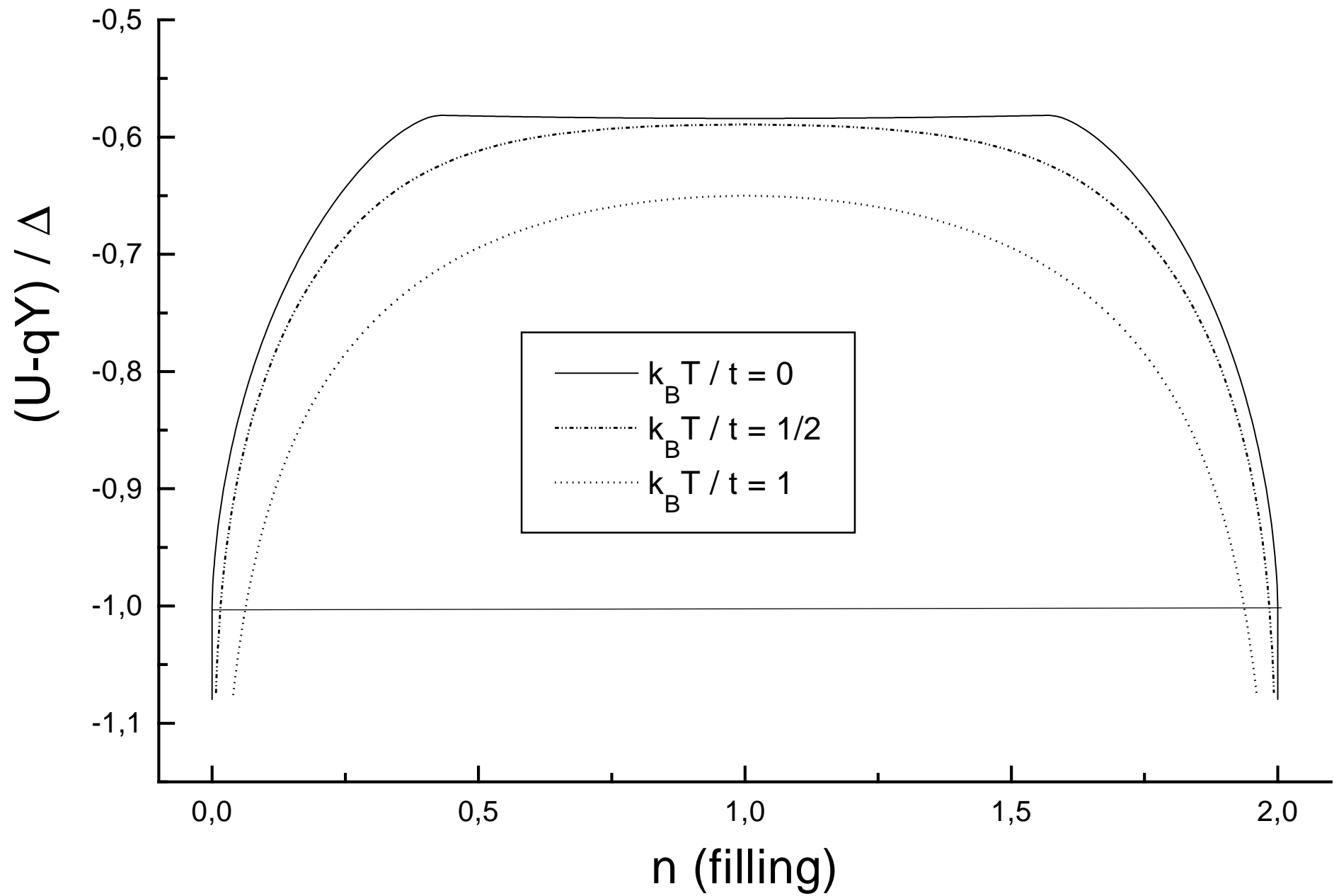
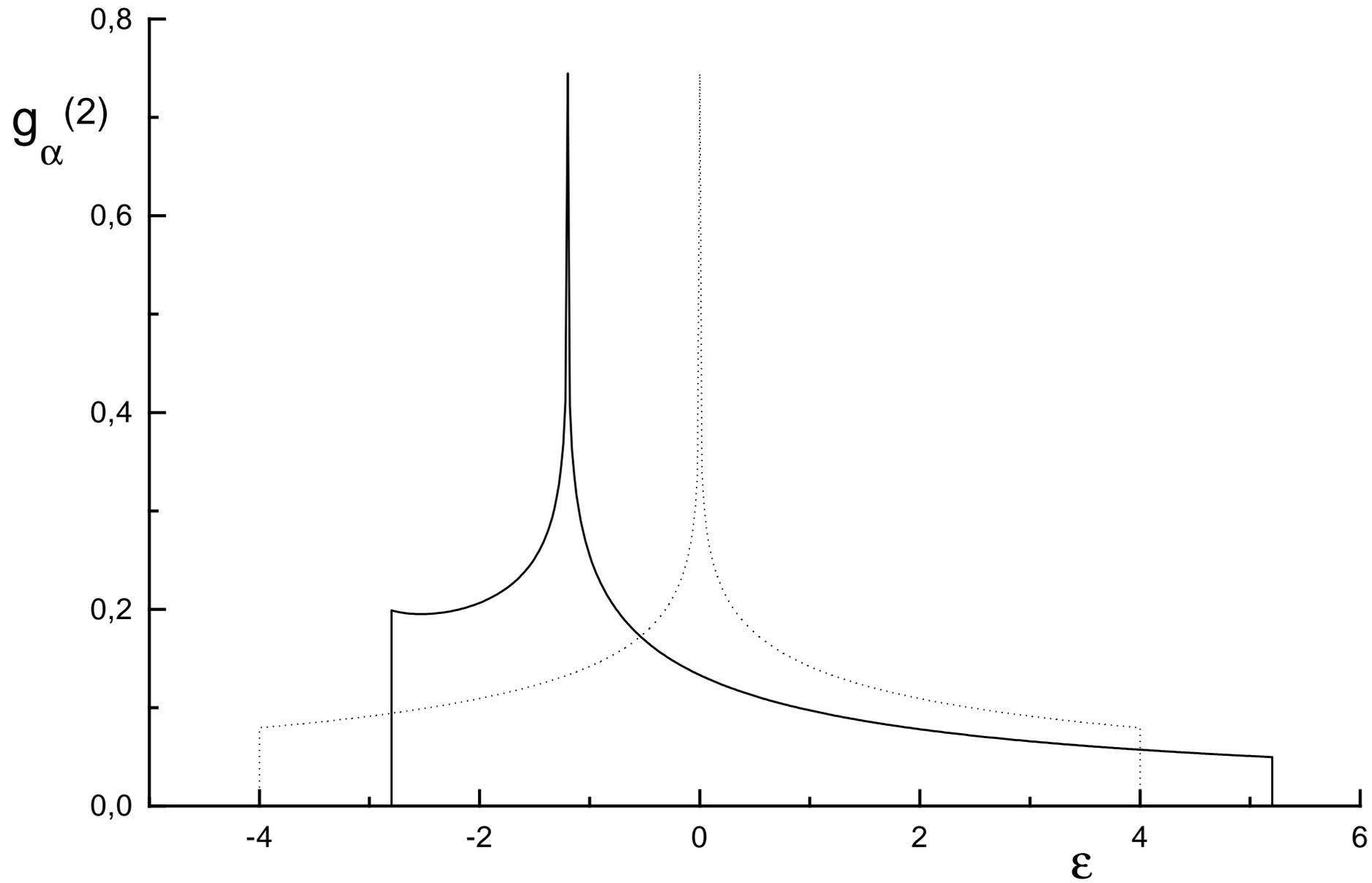
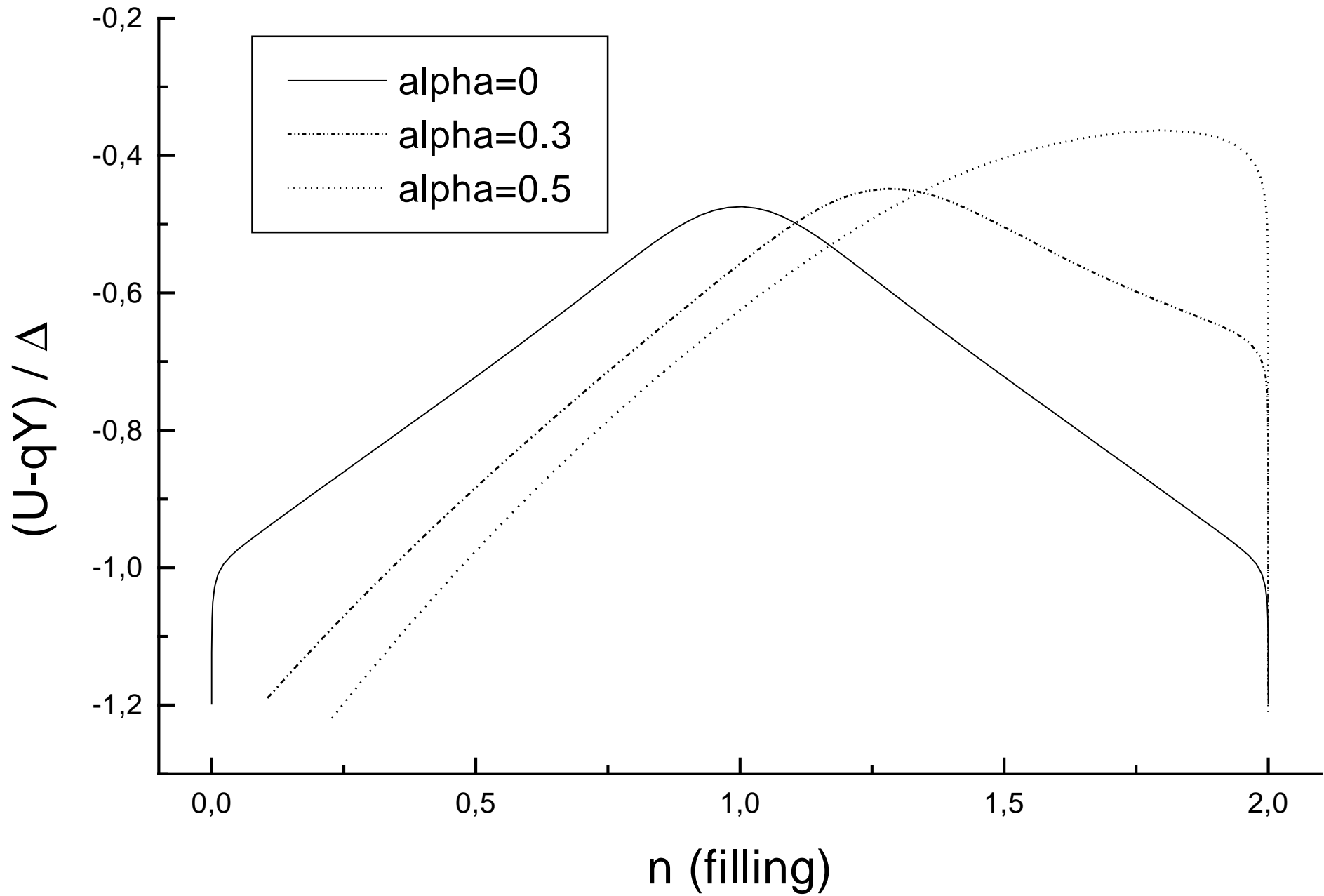
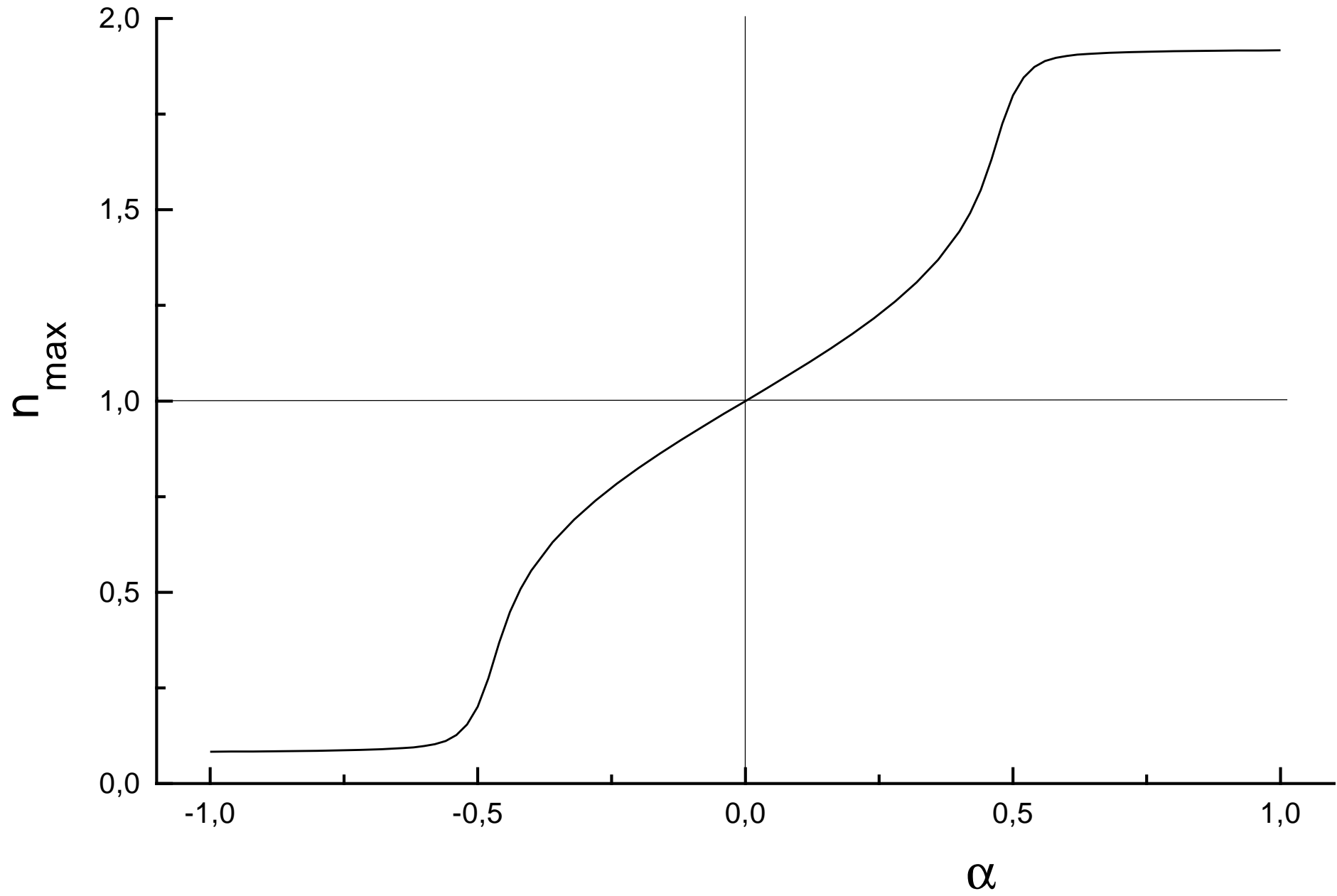
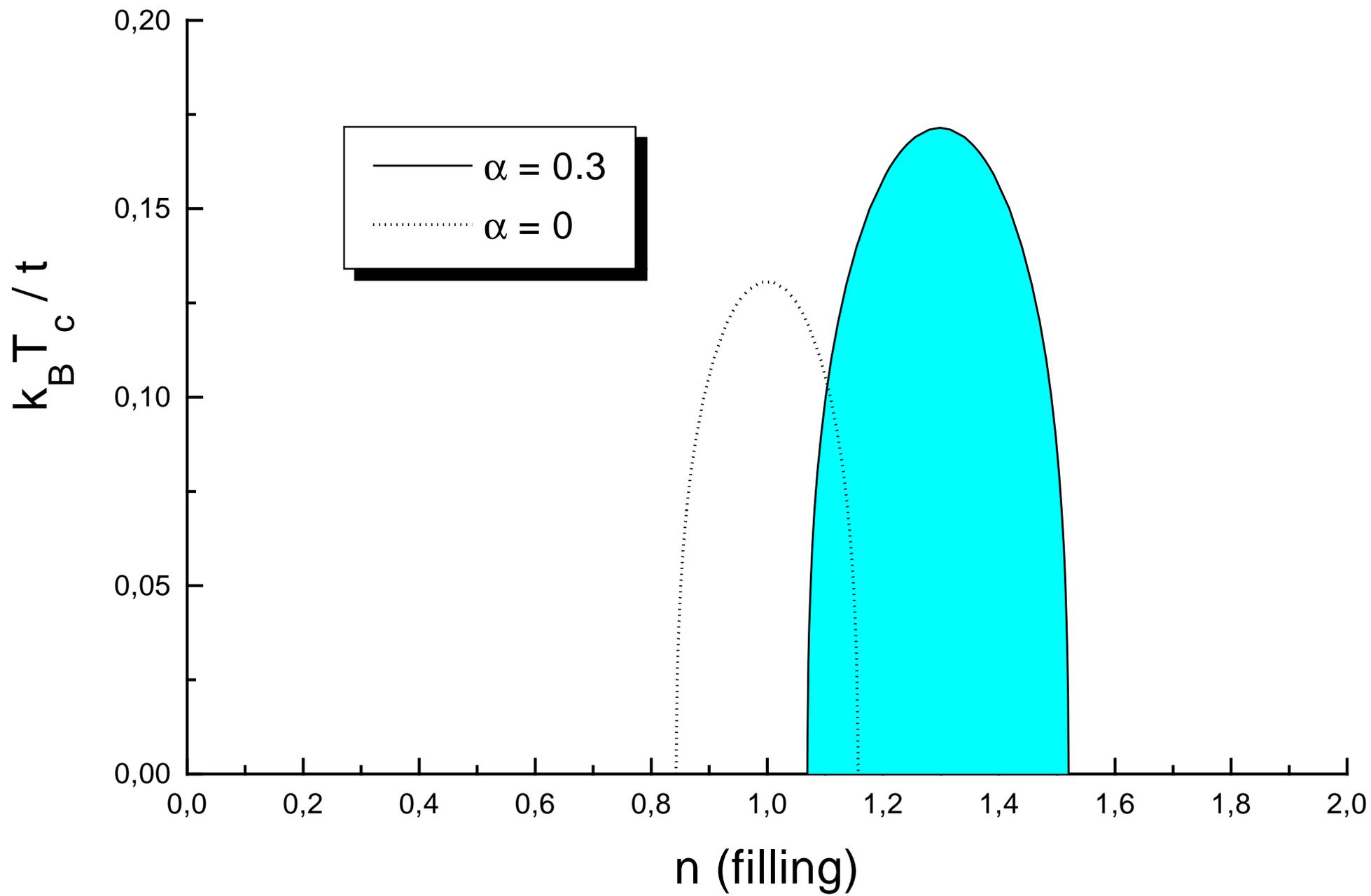
\end{document}